\documentstyle[epsfig]{aipproc}

\begin{document}

\title{A study of hybrid quarkonium using lattice QCD}
\author{K.J.~Juge, J.~Kuti, and C.J.~Morningstar}
\address{Dept.~of Physics, University of California at San Diego,
        La Jolla, California 92093-0319}

\maketitle

\begin{abstract}
The hybrid quarkonium states are studied using the Born-Oppenheimer
expansion.  The first step in this expansion is the determination of
the energy levels of the gluons in the presence of a static
quark-antiquark pair as a function of the quark-antiquark separation.
The spectrum of such gluonic excitations is determined from first
principles using lattice QCD.
\end{abstract}

\section*{Introduction}
In addition to conventional hadrons, QCD suggests the existence of states
containing excited glue, such as glueballs and hybrid hadrons.  Although
conventional hadrons are reasonably well described by the constituent
quark model, states comprised of gluonic excitations are still poorly
understood.  As experiments begin to focus on the search for glueballs
and hybrid mesons, a better understanding of these states is needed.
Lattice QCD simulations presently offer the best means of achieving
this goal.

A great advantage in studying hybrid mesons comprised of heavy quarks
is that such systems can be studied not only by direct numerical
simulation, but also using the Born-Oppenheimer (BO) expansion.  In this
approach, the hybrid meson is treated analogous to a diatomic molecule:
the slow heavy quarks correspond to the nuclei and the fast gluon
field corresponds to the electrons[1].
First, one treats the quark $Q$ and antiquark $\overline{Q}$ as
spatially-fixed colour sources and determines the
energy levels of the glue as a function of the $Q\overline{Q}$ separation
$r$; each of these energy levels defines an adiabatic potential
$V_{Q\overline{Q}}(r)$.  The quark motion is then restored
by solving the Schr\"odinger equation in each of these potentials.
Conventional quarkonia are based on the lowest-lying static potential;
hybrid quarkonium states emerge from the excited potentials.  Once the
static potentials have been determined (via lattice simulations),
it is a simple matter to determine the complete conventional and hybrid
quarkonium spectrum in the leading Born-Oppenheimer (LBO) approximation.
This is a distinct advantage over meson simulations which yield only
the very lowest-lying states, often with large statistical uncertainties.

Here, we present results for the spectrum of gluonic excitations in the
presence of a static quark-antiquark pair.  This study is the
first to comprehensively survey the spectrum in SU(3) gauge theory.
Using our potentials, we also determine the hybrid quarkonium spectrum. 

\section*{Hybrid quarkonium}

The first step in the Born-Oppenheimer expansion is the determination
of the energy levels of the gluons in the presence
of the quark and antiquark, fixed in space some distance $r$ apart.  At
this point in the approximation, the quark and antiquark simply act as
static colour sources.  The gluonic energies (or static potentials)
may be labelled by the magnitude (denoted by $\Lambda$) of the projection
of the total angular momentum of the gluons onto the molecular axis,
by the sign of this projection (chirality or handedness), and by
the behaviour under the combined operations of charge conjugation
and spatial inversion about the midpoint between the quark and the
antiquark.  States with $\Lambda=0,1,2,\dots$
are typically denoted by the capital Greek letters $\Sigma, \Pi,
\Delta, \dots$, respectively.  States which are even (odd) under
the above-mentioned parity--charge-conjugation operation are denoted
by the subscripts $g$ ($u$).  The energy of the gluons is unaffected
by reflections in a plane containing the molecular axis; since such
a reflection interchanges states of opposite handedness, such states
must necessarily be degenerate ($\Lambda$ doubling).  However, this
doubling does not apply to the $\Sigma$ states; $\Sigma$ states which
are even (odd) under a reflection in a plane containing the molecular
axis are denoted by a superscript $+$ $(-)$.  Hence, the low-lying
levels are labelled $\Sigma_g^+$, $\Sigma_g^-$, $\Sigma_u^+$, $\Sigma_u^-$,
$\Pi_g$, $\Pi_u$, $\Delta_g$, $\Delta_u$, and so on.

\begin{figure}[t]
\begin{center}
\leavevmode
\epsfxsize=4.0in\epsfbox[18 164 592 690]{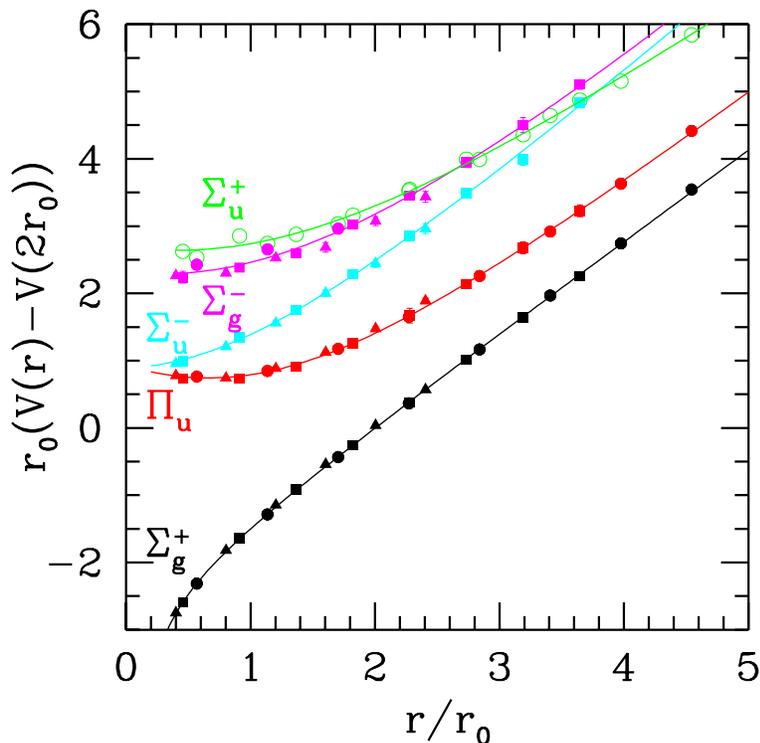}
\end{center}
\caption[figsig]{The static quark potential $V_{\Sigma_g^+}(r)$ and
 some of its gluonic excitations in terms of the hadronic scale
 parameter $r_0$ against the quark-antiquark separation $r$.}
\label{fig:sigma}
\end{figure}

Static potentials were extracted from Monte Carlo estimates 
of generalized Wilson loops; some of our results are shown
in Fig.~\ref{fig:sigma} (for a complete presentation of our results,
see Ref.~[2]).  The results are expressed in terms of
the hadronic scale parameter $r_0$.  The definition of this parameter
and a description of its calculation are given in Ref.~[3].
The familiar static potential is shown as $\Sigma_g^+$.  For all $r$
studied, the first-excited potential is the $\Pi_u$; hence, the lowest
lying hybrid mesons should be based on this potential.

As $r$ becomes very large, the linearly rising $\Sigma_g^+$ potential
suggests that the ground state of the glue may be modelled as a fluctuating
tube of colour flux; in this picture, the gluonic excitations
are expected to be phonon-like with energy gaps proportional to $1/r$.
However, it appears that for $r$ below about $1.5$ fm, the gluonic spectrum
cannot be explained in terms of a simple string model.
In Ref.~[1], a QCD motivated bag model was successfully used
to describe both the $\Sigma_g^+$ and $\Pi_u$ potentials for this range
of $r$.  In this picture, the strong chromoelectric fields of the quark
and antiquark repel the physical vacuum (dual Meissner
effect), creating a bubble inside which perturbation theory is applicable.
In the ground state, the inward pressure on the bubble from the physical
vacuum balances the outward chromostatic force in such a way to produce
a linearly confining potential.  The addition of one or more 
transverse gluons into the
bag produces the excited potentials; the kinetic energy of the gluons
inside the bubble is a key factor in determining the form of these potentials.
This model has recently been revisited and results (in the ellipsoidal
approximation) for almost all of the potentials studied here are in
remarkable agreement with our findings from the lattice simulations
(see Ref.~[4]).

\begin{figure}[t]
\begin{center}
\leavevmode
\epsfxsize=4.0in\epsfbox[18 164 550 680]{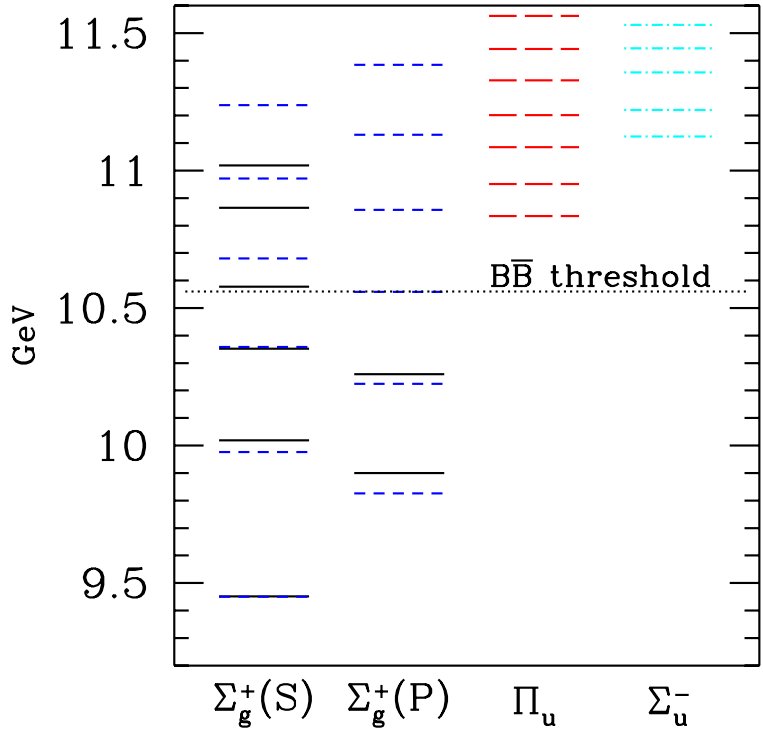}
\end{center}
\caption[figspectrum]{Spin-averaged $b\bar b$ spectrum
 in the leading Born-Oppenheimer and quenched approximations.  Solid lines
 indicate experimental measurements.  Short dashed lines indicate the $S$ and
 $P$ state masses obtained by solving the appropriate Schr\"odinger
 equation in the $\Sigma_g^+$ potential using $r_0^{-1}=0.430$ GeV
 and $M_b=4.60$ GeV for the heavy quark mass. Long dashed and dashed-dotted
 lines indicate the hybrid quarkonium states obtained from the $\Pi_u$
 and $\Sigma_u^-$ potentials, respectively.
\label{fig:spectrum}}
\end{figure}

The next step in the BO expansion is to restore the quark motion by
solving the radial Schr\"odinger equation using the static potentials.
Results for the $b\bar b$ spectrum are shown in~Fig.~\ref{fig:spectrum}.
The heavy quark mass $M_b$ was tuned in order to reproduce the
experimentally-known $\Upsilon(1S)$ mass: $M_\Upsilon
=2M_b+E_0$, where $E_0$ is the energy of the lowest-lying state in the
$\Sigma_g^+$ potential.  In the LBO approximation, many
mesons of different $J^{PC}$ are degenerate, such as
$0^{-+}$,$0^{+-}$,$1^{++}$,$1^{-+}$,$1^{--}$,$1^{+-}$ from the $\Pi_u$
potential.  Below the $B\overline{B}$ threshold, the
LBO results are in very good agreement with the
spin-averaged experimental measurements.  Note that these results make
use of the quenched potentials (which ignore the light quarks) and do not
include spin, retardation, and other relativistic effects.  Above the
threshold, agreement with experiment is lost, suggesting significant
corrections from either the light quarks, relativistic effects, or possibly
mixings between the states from the different adiabatic potentials.  Note
that the mass of the lowest-lying hybrid (from the $\Pi_u$ potential) is
about 10.8~GeV. Above 11 GeV, the LBO approximation based
on the quenched $V_{Q\overline{Q}}$ potentials predicts a very dense
population of hybrid states.

\section*{Conclusion}

A first comprehensive survey of the spectrum of quenched SU(3) gluonic
excitations in the presence of a static $Q\overline{Q}$ pair was presented.
The hybrid quarkonium states were calculated in the leading Born-Oppenheimer
approximation.  This work was supported by the U.S.~DOE,
Grant No.\ DE-FG03-90ER40546.

\end{document}